\documentstyle[12pt,epsf]{article}
\input epsf.tex

\begin{document}
\hoffset = -1.4truecm \voffset = -1.5truecm
\newcommand{\beq}{\begin{equation}}
\newcommand{\eeq}{\end{equation}}
\newcommand{\beqa}{\begin{eqnarray}}
\newcommand{\eeqa}{\end{eqnarray}}
\newcommand{\beqar}{\begin{eqnarray*}}
\title{{\bf Polarized Neutron Matter:\\ A Lowest Order Constrained Variational Approach}}

\author{{\bf G.H. Bordbar \footnote{Corresponding author} \footnote{E-mail :
bordbar@physics.susc.ac.ir}} and {\bf M. Bigdeli}\\
Department of Physics, Shiraz University,
Shiraz 71454, Iran\footnote{Permanent address}\\
and\\
Research Institute for Astronomy and Astrophysics of Maragha,\\
P.O. Box 55134-441, Maragha, Iran }


\maketitle


\begin{abstract}
In this paper, we calculate some of the polarized neutron matter
properties, using the lowest order constrained variational method
with the $AV_{18}$ potential and employing a microscopic point of
view. A comparison is also made between our results and those of
other many-body techniques.
\end{abstract}
21.65.+f, 26.60.+c, 64.70.-p

\section{Introduction}
Pulsars are rapidly rotating neutron stars with strong surface
magnetic fields in the range of $10^{12} -10^{13}$ Gauss
\cite{shap,paci,gold}. The physical origin of this magnetic field
remains an open problem and there is still no general consensus
regarding the mechanism to generate such strong magnetic fields in
a neutron star. There exist several possibilities of the
generation of the magnetic field in a neutron star, from the
nuclear physics point of view, however, one of the most
interesting and stimulating mechanisms which have been suggested
is the possible existence of a phase transition to a ferromagnetic
state at densities corresponding to the theoretically stable
neutron stars and, therefore, of a ferromagnetic core in the
liquid interior of such compact objects. Such a possibility has
been studied by several authors using different theoretical
approaches [4-25], but the results are still contradictory.
Whereas some calculations, like for instance the ones based on
Skyrmelike interactions predict the transition to occur at
densities in the range $(1-4)\rho_{0}$ ($\rho_{0} = 0.16
fm^{-3}$), others, like recent Monte Carlo \cite{fanto} and
Brueckner-Hartree-Fock calculations [21-23] using modern two- and
three-body realistic interactions exclude such a transition, at
least up to densities around five times $\rho_{0}$. This
transition could have important consequences for the evolution of
a protoneutron star, in particular for the spin correlations in
the medium which do strongly affect the neutrino cross section and
the neutrino mean free path inside the star \cite{navarro}.

In recent years, we have computed the equation of state of
symmetrical and asymmetrical nuclear matter and some of their
properties such as symmetry energy, pressure, etc. [27-30] and
properties of spin polarized liquid $^{3}He$ \cite{bord05} using
the lowest order constrained variational (LOCV) approach. The LOCV
method which was developed several years ago is a useful tool for
the determination of the properties of neutron, nuclear and
asymmetric nuclear matter at zero and finite temperature [27-39].
The LOCV method is a fully self-consistent formalism and it does
not bring any free parameters into calculation. It employs a
normalization constraint to keep the higher order term as small as
possible \cite{owen,borda}. The functional minimization procedure
represents an enormous computational simplification over
unconstrained methods that attempt to go beyond lowest order.

In the present work, we compute the polarized neutron matter
properties using the LOCV method with the $AV_{18}$ potential
\cite{wiring} employing microscopic calculations where we treat
explicitly the spin projection in the many-body wave functions.

\section{Basic Theory}

\subsection{LOCV Formalism}
We consider a trial many-body wave function of the form
\begin{eqnarray}
     \psi=F\phi,
 \end{eqnarray}
where $\phi$ is the uncorrelated ground state wave function
(simply the Slater determinant of plane waves) of $N$ independent
neutron and $F=F(1\cdots N)$ is an appropriate N-body correlation
operator which can be replaced by a Jastrow form i.e.,
\begin{eqnarray}
    F=S\prod _{i>j}f(ij),
 \end{eqnarray}
in which S is a symmetrizing operator. We consider a cluster
expansion of the energy functional up to the two-body term,
 \begin{eqnarray}
           E([f])=\frac{1}{N}\frac{\langle\psi|H\psi\rangle}
           {\langle\psi|\psi\rangle}=E _{1}+E _{2}\cdot
 \end{eqnarray}
The one-body term $E _{1}$ for a polarized neutron matter can be
written as Fermi momentum functional ($k
_{F}^{(i)}=(6\pi^{2}\rho^{(i)})^{\frac{1}{3}})$:
\begin{eqnarray}\label{ener1}
               E _{1}=\sum _{i=1,2}\frac{3}{5}
               \frac{\hbar^{2}{k
               _{F}^{(i)}}^2}{2m}\frac{\rho^{(i)}}{\rho}\cdot
 \end{eqnarray}
Labels 1 and 2 are used instead of spin up and spin down neutrons,
respectively, and $\rho=\rho^{(1)}+\rho^{(2)}$ is the total
neutron matter density. The two-body energy $E_{2}$ is
\begin{eqnarray}
    E_{2}&=&\frac{1}{2A}\sum_{ij} \langle ij\left| \nu(12)\right|
    ij-ji\rangle,
 \end{eqnarray}
where \\ $\nu(12)=-\frac{\hbar^{2}}{2m}[f(12),[\nabla
_{12}^{2},f(12)]]+f(12)V(12)f(12)$, $f(12)$ and $V(12)$ are the
two-body correlation and potential. For the two-body correlation
function, $f(12)$, we consider the following form
\cite{borda,bordb}:
\begin{eqnarray}
f(12)&=&\sum^3_{k=1}f^{(k)}(12)O^{(k)}(12),
\end{eqnarray}
where, the operators $O^{(k)}(12)$ are given by
\begin{eqnarray}
O^{(k=1-3)}(12)&=&1,\ (\frac{2}{3}+\frac{1}{6}S_{12}),\
(\frac{1}{3}-\frac{1}{6}S_{12}),
\end{eqnarray}
and $S_{12}$ is the tensor operator.

After doing some algebra we find the following equation for the
two-body energy:
\begin{eqnarray}\label{ener2}
    E_{2} &=& \frac{2}{\pi ^{4}\rho }\left( \frac{h^{2}}{2m}\right)
    \sum_{JLSS_{z}}\frac{(2J+1)}{2(2S+1)}[1-(-1)^{L+S+1}]\left| \left\langle
\frac{1}{2}\sigma _{z1}\frac{1}{2}\sigma _{z2}\mid
SS_{z}\right\rangle \right| ^{2} \int dr\left\{\left [{f_{\alpha
}^{(1)^{^{\prime }}}}^{2}{a_{\alpha
}^{(1)}}^{2}(k_{f}r)\right.\right. \nonumber
\\&& \left.\left. +\frac{2m}{h^{2}}(\{V_{c}-3V_{\sigma } +V_{\tau }-3V_{\sigma
\tau }+2(V_{T}-3V_{\sigma \tau }) +2V_{\tau z}\}{a_{\alpha
}^{(1)}}^{2}(k_{f}r)\right.\right. \nonumber \\&&\left.\left.
+[V_{l2}-3V_{l2\sigma } +V_{l2\tau }-3V_{l2\sigma \tau
}]{c_{\alpha }^{(1)}}^{2}(k_{f}r))(f_{\alpha }^{(1)})^{2}\right ]
+\sum_{k=2,3}\left[ {f_{\alpha }^{(k)^{^{\prime }}}}^{2}{a_{\alpha
}^{(k)}}^{2}(k_{f}r)\right.\right. \nonumber \\&&\left. \left.
+\frac{2m}{h^{2}}( \{V_{c}+V_{\sigma }+V_{\tau } +V_{\sigma \tau
}+(-6k+14)(V_{tz}+V_{t})-(k-1)(V_{ls\tau }+V_{ls})\right.\right.
\nonumber
\\&&\left.\left. +[V_{T}+V_{\sigma \tau }+(-6k+14)V_{tT}] [2+2V_{\tau
z}]\}{a_{\alpha }^{(k)}}^{2}(k_{f}r)\right.\right. \nonumber
\\&&\left.\left. +[V_{l2}+V_{l2\sigma } +V_{l2\tau }+V_{l2\sigma \tau
}]{c_{\alpha }^{(k)}}^{2}(k_{f}r)+[V_{ls2}+V_{ls2\tau }]
{d_{\alpha }^{(k)}}^{2}(k_{f}r)) {f_{\alpha }^{(k)}}^{2}\right
]\right. \nonumber \\&&\left. +\frac{2m}{h^{2}}\{V_{ls}+V_{ls\tau
}-2(V_{l2}+V_{l2\sigma }+V_{l2\sigma \tau } +V_{l2\tau
})-3(V_{ls2} +V_{ls2\tau })\}b_{\alpha }^{2}(k_{f}r)f_{\alpha
}^{(2)}f_{\alpha }^{(3)}\right. \nonumber \\&&\left.
+\frac{1}{r^{2}}(f_{\alpha }^{(2)} -f_{\alpha
}^{(3)})^{2}b_{\alpha }^{2}(k_{f}r)\right\},
 \end{eqnarray}
where $\alpha=\{J,L,S,S_z\}$ and the coefficient  ${a_{\alpha
}^{(1)}}^{2}$, etc., are defined as
\begin{eqnarray}
     {a_{\alpha }^{(1)}}^{2}(x)=x^{2}I_{L,S_{z}}(x),
 \end{eqnarray}
\begin{eqnarray}
     {a_{\alpha }^{(2)}}^{2}(x)=x^{2}[\beta I_{J-1,S_{z}}(x)
     +\gamma I_{J+1,S_{z}}(x)],
 \end{eqnarray}
\begin{eqnarray}
           {a_{\alpha }^{(3)}}^{2}(x)=x^{2}[\gamma I_{J-1,S_{z}}(x)
           +\beta I_{J+1,S_{z}}(x)],
      \end{eqnarray}
\begin{eqnarray}
     b_{\alpha }^{(2)}(x)=x^{2}[\beta _{23}I_{J-1,S_{z}}(x)
     -\beta _{23}I_{J+1,S_{z}}(x)],
 \end{eqnarray}
\begin{eqnarray}
         {c_{\alpha }^{(1)}}^{2}(x)=x^{2}\nu _{1}I_{L,S_{z}}(x),
      \end{eqnarray}
\begin{eqnarray}
        {c_{\alpha }^{(2)}}^{2}(x)=x^{2}[\eta _{2}I_{J-1,S_{z}}(x)
        +\nu _{2}I_{J+1,S_{z}}(x)],
 \end{eqnarray}
\begin{eqnarray}
       {c_{\alpha }^{(3)}}^{2}(x)=x^{2}[\eta _{3}I_{J-1,S_{z}}(x)
       +\nu _{3}I_{J+1,S_{z}}(x)],
 \end{eqnarray}
\begin{eqnarray}
     {d_{\alpha }^{(2)}}^{2}(x)=x^{2}[\xi _{2}I_{J-1,S_{z}}(x)
     +\lambda _{2}I_{J+1,S_{z}}(x)],
 \end{eqnarray}
\begin{eqnarray}
     {d_{\alpha }^{(3)}}^{2}(x)=x^{2}[\xi _{3}I_{J-1,S_{z}}(x)
     +\lambda _{3}I_{J+1,S_{z}}(x)],
 \end{eqnarray}
with
\begin{eqnarray}
          \beta =\frac{J+1}{2J+1},\ \gamma =\frac{J}{2J+1},\
          \beta _{23}=\frac{2J(J+1)}{2J+1},
 \end{eqnarray}
\begin{eqnarray}
       \nu _{1}=L(L+1),\ \nu _{2}=\frac{J^{2}(J+1)}{2J+1},\
       \nu _{3}=\frac{J^{3}+2J^{2}+3J+2}{2J+1},
      \end{eqnarray}
\begin{eqnarray}
     \eta _{2}=\frac{J(J^{2}+2J+1)}{2J+1},\ \eta _{3}=
     \frac{J(J^{2}+J+2)}{2J+1},
 \end{eqnarray}
\begin{eqnarray}
     \xi _{2}=\frac{J^{3}+2J^{2}+2J+1}{2J+1},\
     \xi _{3}=\frac{J(J^{2}+J+4)}{2J+1},
 \end{eqnarray}
\begin{eqnarray}
     \lambda _{2}=\frac{J(J^{2}+J+1)}{2J+1},\
     \lambda _{3}=\frac{J^{3}+2J^{2}+5J+4}{2J+1},
 \end{eqnarray}

and
\begin{eqnarray}
       I_{J,S_{z}}(x)=\int dqP_{S_{z}}(q)J_{J}^{2}(xq)\cdot
 \end{eqnarray}
In the above equation $J_{J}(x)$ is the Bessel's function and,
$P_{S_{z}}(q)$ is defined as follows,
\begin{eqnarray}
       P_{S_{z}}(q)&=&\frac{2}{3}\pi \lbrack (k_{F}^{\sigma _{z1}})^{3}
       +(k_{F}^{\sigma _{z2}})^{3}
       -\frac{3}{2}((k_{F}^{\sigma _{z1}})^{2} +(k_{F}^{\sigma _{z2}})^{2})q
       \nonumber\\&&
       -\frac{3}{16}((k_{F}^{\sigma _{z1}})^{2}
       -(k_{F}^{\sigma _{z2}})^{2})^{2}q^{-1}+q^{3}],
 \end{eqnarray}
for $\frac{1}{2}\left| k_{F}^{\sigma _{z1}}-k_{F}^{\sigma
_{z2}}\right|
        < q< \frac{1}{2}\left| k_{F}^{\sigma _{z1}}+k_{F}^{\sigma
        _{z2}}\right|$,
\begin{eqnarray}
      P_{S_{z}}(q)=\frac{4}{3}\pi min(k_{F}^{\sigma _{z1}},k_{F}^{\sigma
      _{z2}}),
 \end{eqnarray}
 for $q<\frac{1}{2}\left| k_{F}^{\sigma _{z1}}-k_{F}^{\sigma
 _{z2}}\right|$ and
\begin{eqnarray}
            P_{S_{z}}(q)=0,
 \end{eqnarray}

for $q>\frac{1}{2}\left| k_{F}^{\sigma _{z1}}+k_{F}^{\sigma
 _{z2}}\right|$,
 where $\sigma _{z1}$ or $\sigma _{z2}=\frac{1}{2},-\frac{1}{2}$
for spin up and spin down, respectively.

Now, we can minimize the two-body energy Eq.(\ref{ener2}), with
respect to the variations in the function ${f_{\alpha}}^{(i)}$ but
subject to the normalization constraint \cite{bordb},

\begin{eqnarray}
        \frac{1}{A}\sum_{ij}\langle ij\left| h_{S_{z}}^{2}
        -f^{2}(12)\right| ij\rangle _{a}=0,
 \end{eqnarray}
where in the case of spin polarized neutron matter the function
$h_{S_{z}}(r)$ is defined as
\begin{eqnarray}
       h_{S_{z}}(r)&=&\left[ 1-9\left( \frac{J_{J}^{2}
       (k_{F}^{i})}{k_{F}^{i}}\right) ^{2}\right] ^{-1/2};\  S_{z}=\pm1
       \nonumber\\
       &=& 1\ \ \ \ \ \ \ \ \ \ \ \ \ \ \ \ \ \ \ \ \ \ \ \ \ \ \
       \ \ \ ;\ S_{z}= 0
 \end{eqnarray}
From the minimization of the two-body cluster energy, we get a set
of coupled and uncoupled
 differential equations which are the same as presented in Ref.\cite{bordb}.

\subsection{Magnetic Susceptibility}

The magnetic susceptibility, $\chi$, which characterizes the
response of a system to the magnetic field, is defined by
\begin{eqnarray}
\chi =\left( \frac{\partial M}{\partial H}\right) _{H=0},
\end{eqnarray}
where $M$ is the magnetization of system per unit volume and $H$
is the magnetic field. By some simplification, the magnetic
susceptibility can be written as
\begin{eqnarray}\label{susep}
   \chi =\frac{\mu ^{2}\rho
}{\left( \frac{\partial ^{2} E }{\partial \delta ^{2}}\right)
_{\delta =0}},
\end{eqnarray}
where $\mu$ is the magnetic moment of the neutron and $\delta$ is
the spin polarization parameter which is defined as
\begin{eqnarray}\label{pol}
\delta =\frac{\rho ^{(1)}-\rho ^{(2)}}{\rho }\cdot
\end{eqnarray}
Usually, one is interested in calculating the ratio of $\chi$ to
the magnetic susceptibility for a degenerate free Fermi gas
($\chi_{F}$). $\chi_{F}$ can be straightforwardly obtained from
Eq. (\ref{susep}), using the total energy per particle of free
Fermi gas,
\begin{eqnarray}
 \chi_{F} =\frac{\mu
^{2}m}{\hbar ^{2}\pi ^{2}}k_{F},
\end{eqnarray}
where $k_{F}=(3\pi^{2}\rho)^{1/3}$ is Fermi momentum. After a
little algebra one finds
\begin{eqnarray}
   \frac{\chi}{\chi_{F}} =\frac{2}{3}\frac{E_{F}
}{\left( \frac{\partial ^{2} E }{\partial \delta ^{2}}\right)
_{\delta =0}},
\end{eqnarray}
where $E_{F}={\hbar ^{2}k_{F}^{2}}/{2m}$ is the Fermi energy.
%
\section{Results}\label{NLmatchingFFtex}

In Fig. 1, we have shown the energy per particle for various
values of spin polarization of the neutron matter as a function of
density. As can be seen from this figure, the energy of neutron
matter becomes repulsive by increasing the polarization for all
relevant densities. According to this result, the spontaneous
phase transition to a ferromagnetic state in the neutron matter
does not occur. If such a transition existed a crossing of the
energies of different polarizations would be observed at some
density, indicating that the ground state of the system would be
ferromagnetic from that density on. As is shown in Fig. 1, there
is no crossing point. On the contrary, it becomes less favorable
as the density increases. For the energy of neutron matter, we
have also made a comparison between our results and the results of
other many-body methods with the $AV_{18}$ potential as shown in
Fig. 2. The BGLS  calculations are based on the
Brueckner-Hartree-Fock approximation both for continuous choice
(BHFC) and standard choice (BHFG) \cite{bgls}. The APR results
have been obtained using the variational chain summation (VCS)
method \cite{apr} and the EHMMP calculations have been carried out
using the lowest order Brueckner (LOB) technique \cite{ehmmp}. We
see that our results are in agreement with those of others,
specially with the APR and EHMMP calculations.

For the neutron matter, we have also considered the dependence of
energy to the spin polarization $\delta$. Let us examine this
dependency in quadratic spin polarization form for different
densities as shown in Fig. 3. As can be seen from this figure, the
energy per particle increases as the polarization increases and
the minimum value of energy occurs at $\delta=0$ for all
densities. This indicates that the ground state of neutron matter
is paramagnetic. In Fig. 3, the results of ZLS calculations using
the Brueckner-Hartree-Fock theory with the $AV_{18}$ potential
\cite{zls} are also given for comparison. There is an agreement
between our results and those of ZLS, specially at low densities.
From Fig. 3, it is also seen that the variation of the energy of
neutron matter versus $\delta^{2}$ is nearly linear. Therefore,
one can characterize this dependency in the following analytical
form
\begin{eqnarray}
       E (\rho,\delta)&=&E(\rho,0)+a(\rho)\delta^{2}\cdot
 \end{eqnarray}
The density dependent parameter $a(\rho)$ can be interpreted as
the measure of the energy required to produce a net spin alignment
in the direction of the magnetic field, and its value can be
determined as the slope of each line in Fig. 3, for the
corresponding density,
\begin{eqnarray}
      a(\rho )=\frac{\partial E(\rho,\delta)}{\partial \delta
      ^{2}}\cdot
 \end{eqnarray}
In Fig. 4, the parameter $a(\rho)$ is shown as a function of the
density and as can be seen the value of this parameter increase by
increasing the density. In turn this indicate the energy which
require to align spin at same direction increases. An conclusion
can be inferred again from this result is that a phase transition
to a ferromagnetic state is not to be expected from our
calculation. The parameter $a(\rho)$ obtained by ZLS \cite{zls} is
also shown in Fig. 4, for comparison.

In Fig. 5, we have plotted the ratio ${{\chi}/{\chi_{F}}}$ versus
density. As can be seen from Fig. 5, this ratio changes
continuously for all densities. Therefore, the ferromagnetic phase
transition does not occur. For comparison, we have also shown the
results of ZLS \cite{zls} in this figure.

The equation of state of polarized neutron matter,
$P(\rho,\delta)$, can be simply obtained using

\begin{eqnarray}
      P(\rho,\delta)= \rho^{2} \frac{\partial E(\rho,\delta)}{\partial \rho}
 \end{eqnarray}
In Fig. 6, we have presented the pressure of neutron matter as a
function of density $\rho$ at different polarizations. We see that
the equation of state becomes stiffer by increasing the
polarization. We also see that with increasing density, the
difference between the equations of state at different
polarization becomes more appreciable. In order to check the
causality condition for our equations of state, we have calculated
the velocity of sound, $v_{s}$,  as shown in Fig. 7. It is seen
that the velocity of sound increases with both increasing
polarization and density, but it is always less than the velocity
of light in the vacuum $(c)$. Therefore, all calculated equations
of state obey the causality condition.

As it is known, the Landau parameter, $G_{0}$, describes the spin
density fluctuation in the effective interaction. $G_{0}$ is
simply related to the magnetic susceptibility by the relation
\begin{eqnarray}
   \frac{\chi}{\chi_{F}} =\frac{m^{*}
}{1+G_{0}}
\end{eqnarray}
where $m^{*}$ is the effective mass. A magnetic instability would
require $G_{0}<-1$. Our results for the Landau parameter have been
presented in the Fig. 8. It is seen that the value of $G_{0}$ is
always positive and monotonically increasing up to highest density
and does not show any magnetic instability for the neutron matter.
In Fig. 8, the results of ZLS calculations \cite{zls} are also
given for comparison.


\section{Summary and Conclusions}
The properties of neutron matter is of primary importance in the
study of neutron star, and in particular, strongly magnetized ones
(i.e. pulsars). It is therefore important to calculate the
properties of polarized neutron matter, using an efficient and
sufficiently accurate method. We have recently computed various
properties of the neutron matter using the lowest order
constrained variational (LOCV) scheme. In order to make our
results more general, we used this method for the polarized
neutron matter. Energy per particle for various values of spin
polarization of the neutron matter was computed as function of
density, and shown to become repulsive as a result of increasing
the polarization. In addition, we considered the dependence of
energy of neutron matter to the spin polarization, and found it to
increase with the spin polarization for all densities. This
dependence was represented by a quadratic formula with the
coefficient of the quadratic term $a(\rho)$ determined as a
function of the density. This parameter, too, was shown to
increase monotonically with density. Magnetic susceptibility,
which characterizes the response of the system to the magnetic
field was calculated for the system under consideration. We have
also computed the equation of state of neutron matter at different
polarizations. Our results for higher values of polarization show
a stiff equation of state. The velocity of sound was computed to
check the causality condition of equation of state and it was
shown that it is always lower than the velocity of light in
vacuum. We have also investigated the Landau parameter $G_{0}$
which shows that the value of $G_{0}$ is always positive and
monotonically increasing up to high densities. Finally, our
results showed no phase transition to ferromagnetic state. We have
also compared the results of our calculations for the properties
of neutron matter with the other calculations.

\section*{Acknowledgements} This work has been supported by Research
Institute for Astronomy and Astrophysics of Maragha, and Shiraz
University Research Council.

\newpage

\newpage


\begin{figure}
\centerline{\epsfxsize 4.5 truein \epsfbox {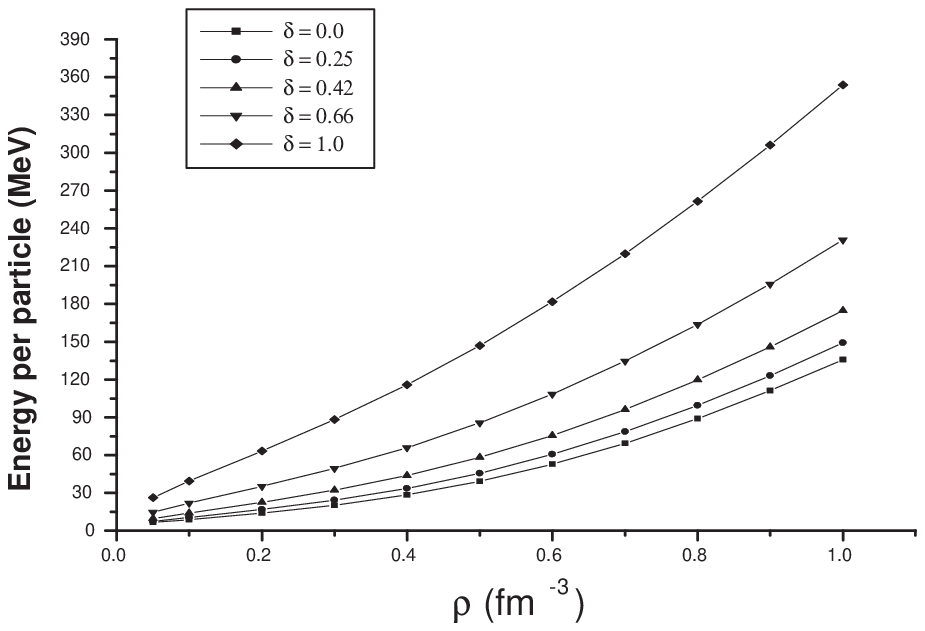}}
\caption{The energy per particle versus density ($\rho$) for
different values of the spin polarization ($\delta$) of the
neutron matter.} \label{EN(den)}
\end{figure}
\newpage

\begin{figure}
\centerline{\epsfxsize 4.5 truein \epsfbox {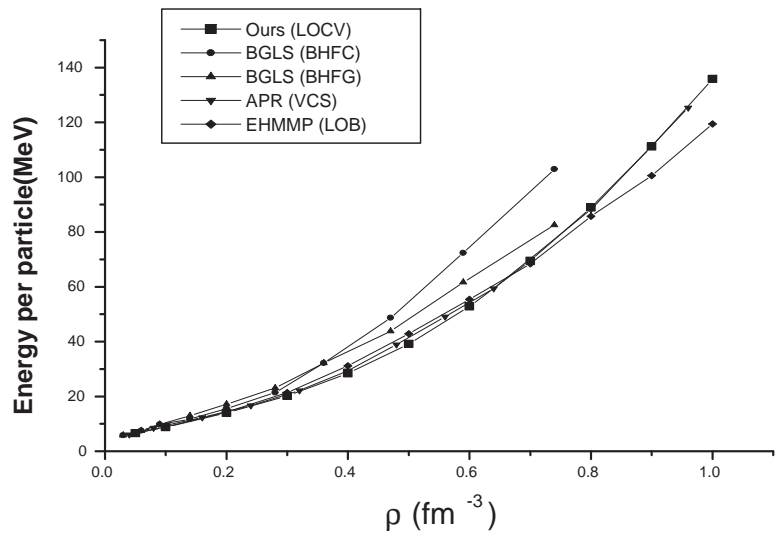}}
\caption{Comparison between our results for the energy per
particle of neutron matter and those of BGLS [41], APR [42] and
EHMMP [43] calculations with the $AV_{18}$ potential.}
\label{EN(den2)}
\end{figure}
\newpage


\begin{figure}

\centerline{\epsfxsize 4.5 truein \epsfbox {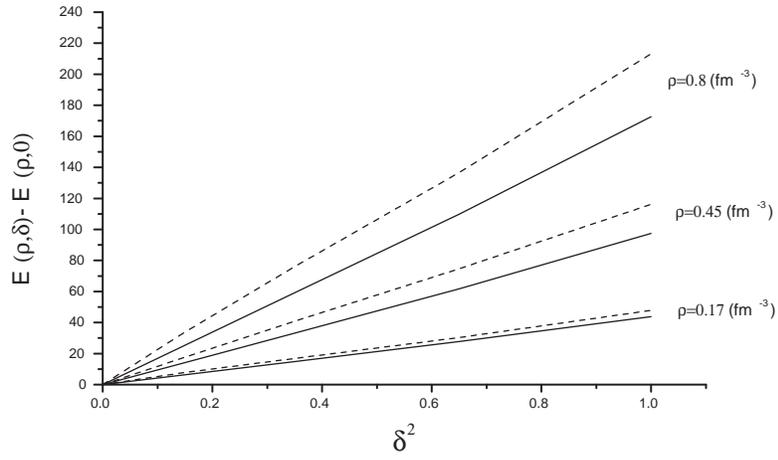}}
\caption{Our results (full curves) for the energy difference of
polarized and unpolarized cases versus quadratic spin polarization
($\delta$) for different values of the density($\rho$) of the
neutron matter. The results of ZLS [25] (dashed curves) are also
presented for comparison.} \label{EN(comp)}
\end{figure}
\newpage


\begin{figure}
\centerline{\epsfxsize 4.5 truein \epsfbox {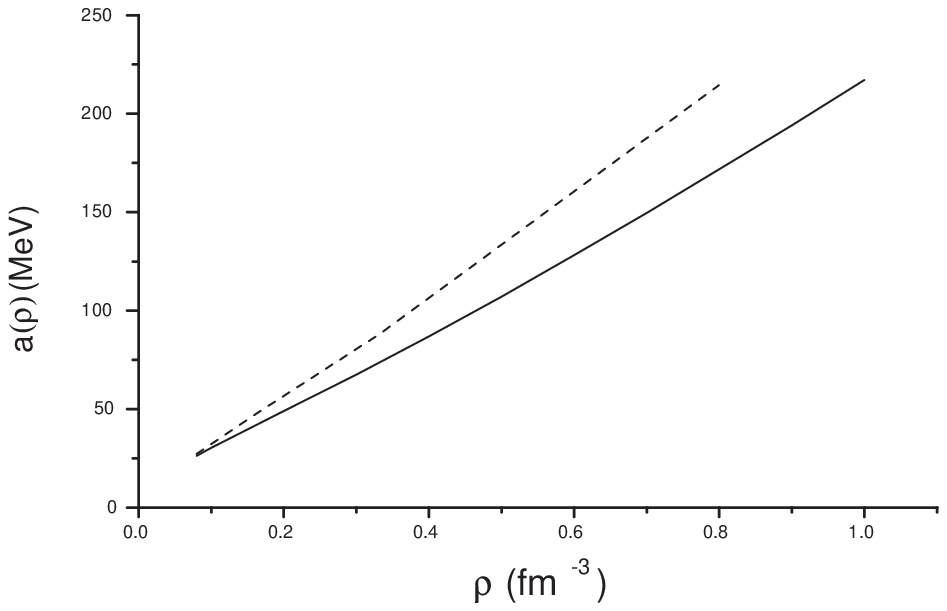}}
\caption{Our results (full curve) for the parameter $a(\rho)$  as
a function of the density($\rho$). The results of ZLS [25] (dashed
curve) are also given for comparison.} \label{a(den)}
\end{figure}

\newpage


\begin{figure} \centerline{\epsfxsize 4.5 truein \epsfbox
{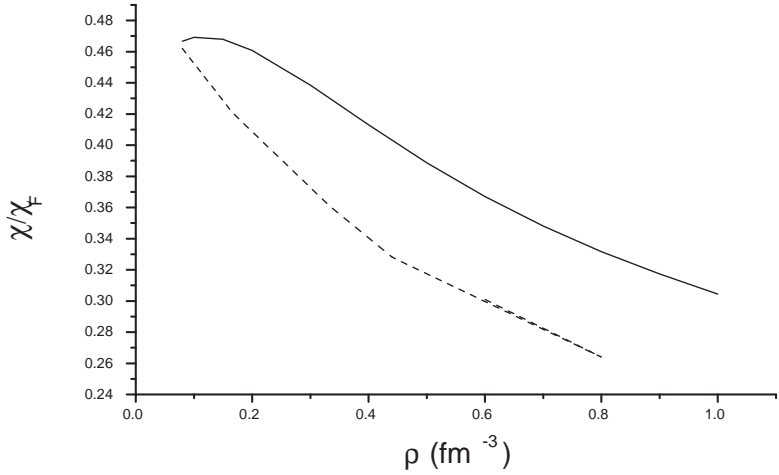}} \caption{As Fig. 4, but for the magnetic
susceptibility (${{\chi}/{\chi_{F}}}$).} \label{sus(den)}
\end{figure}
\newpage


\begin{figure}
\centerline{\epsfxsize 4.5 truein \epsfbox {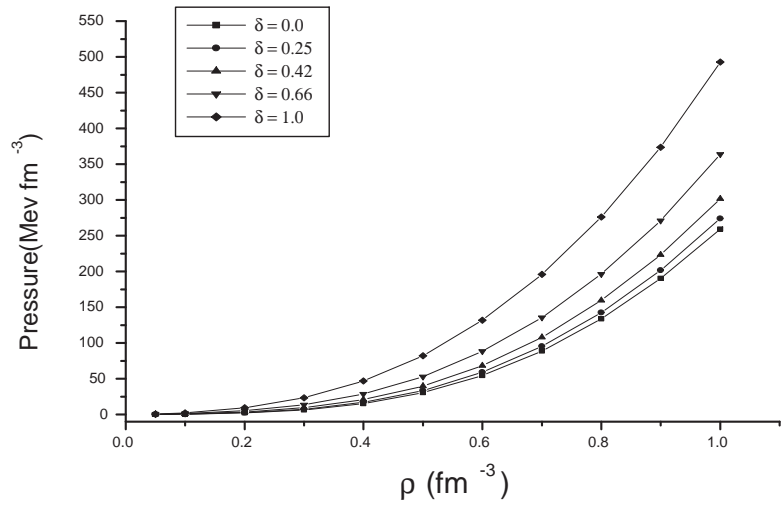}}
\caption{The equation of state of neutron matter for different
values of the spin polarization ($\delta$).} \label{PRE(den)}
\end{figure}
\newpage


\begin{figure}
\centerline{\epsfxsize 4.5 truein \epsfbox {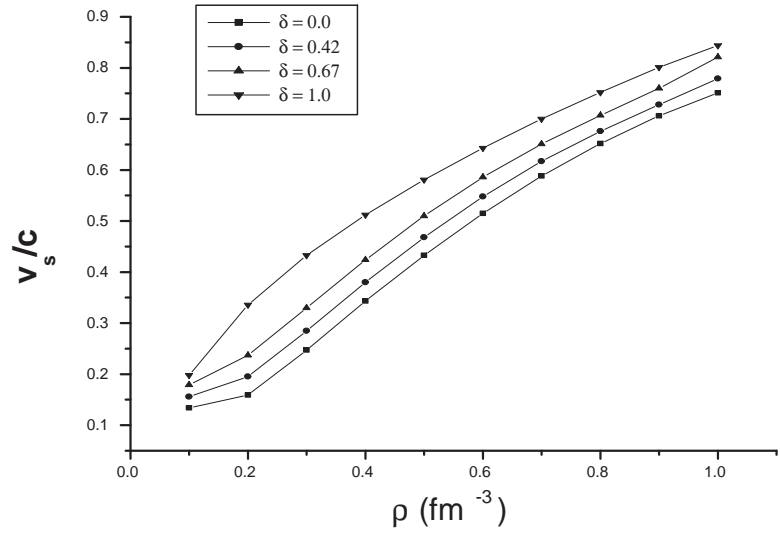}}
\caption{The velocity of sound in the unit of $c$ versus density
($\rho$) for different values of the spin polarization
($\delta$).} \label{VEL(den)}
\end{figure}
\newpage


\begin{figure}
\centerline{\epsfxsize 4.5 truein \epsfbox {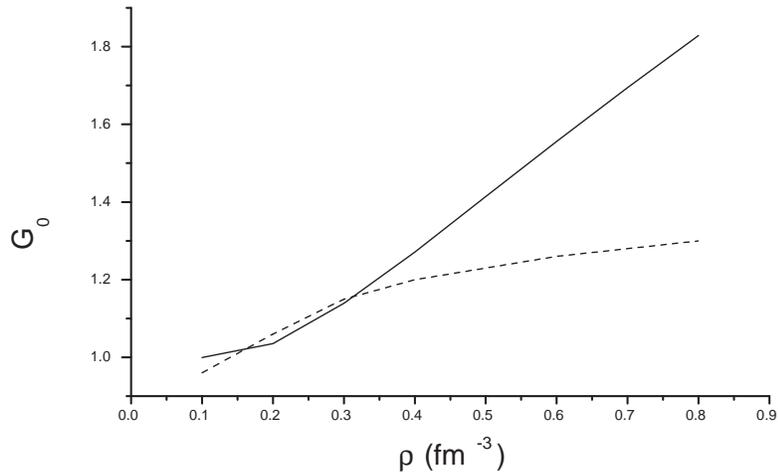}}

\caption{As Fig. 4, but for the Landau parameter, $G_{0}$.}
\label{G(den)}
\end{figure}


\end{document}